\documentclass[journal=nalefd,manuscript=letter]{achemso}
\usepackage{epsf}
\usepackage{graphicx}
\usepackage{dcolumn}
\usepackage{bm}
\usepackage{amsmath}
\usepackage{amsfonts}
\usepackage{amssymb}
\usepackage{ulem}
\usepackage{color}

\title{Controlling Dzyaloshinskii-Moriya Interaction via Chirality Dependent Atomic-Layer Stacking, Insulator Capping and Electric Field}
\author{Hongxin~Yang}
\affiliation{Univ. Grenoble Alpes, INAC-SPINTEC, F-38000 Grenoble, France; CNRS, SPINTEC, F-38000 Grenoble, France; and CEA, INAC-SPINTEC, F-38000 Grenoble, France}
\alsoaffiliation{Unit$\acute{e}$  Mixte de Physique CNRS, Thales, Univ. Paris-Sud, Univ. Paris-Saclay, Palaiseau, France}

\author{Olivier Boulle}
\affiliation{Univ. Grenoble Alpes, INAC-SPINTEC, F-38000 Grenoble, France; CNRS, SPINTEC, F-38000 Grenoble, France; and CEA, INAC-SPINTEC, F-38000 Grenoble, France}

\author{Vincent Cros}
\affiliation{Unit$\acute{e}$  Mixte de Physique CNRS, Thales, Univ. Paris-Sud, Univ. Paris-Saclay, Palaiseau, France}

\author{Albert~Fert}
\affiliation{Unit$\acute{e}$  Mixte de Physique CNRS, Thales, Univ. Paris-Sud, Univ. Paris-Saclay, Palaiseau, France}

\author{Mairbek~Chshiev}
\affiliation{Univ. Grenoble Alpes, INAC-SPINTEC, F-38000 Grenoble, France; CNRS, SPINTEC, F-38000 Grenoble, France; and CEA, INAC-SPINTEC, F-38000 Grenoble, France}

\email{mair.chshiev@cea.fr}

\begin{document}



\begin{abstract}
Using first-principle calculations, we demonstrate several approaches to manipulate Dzyaloshinskii-Moriya Interaction (DMI) in ultrathin magnetic films. First, we find that DMI is significantly enhanced when the ferromagnetic (FM) layer is sandwiched between nonmagnetic (NM) layers inducing additive DMI in NM/FM/NM structures. For instance, as Pt and Ir below Co induce DMI of opposite chirality, inserting Co between Pt (below) and Ir (above) in Ir/Co/Pt trilayers enhances the DMI of Co/Pt bilayers by 15\%. Furthermore, in case of Pb/Co/Pt trilayers (Ir/Fe/Co/Pt multilayers), DMI can be enhanced by 50\% (almost doubled) compared to Co/Pt bilayers reaching a very large DMI amplitude of 2.7 (3.2) meV/atom. Our second approach for enhancing DMI is to use oxide capping layer. We show that DMI is enhanced by 60\% in Oxide/Co/Pt structures compared to Co/Pt bilayers. Moreover, we unveiled the DMI mechanism at Oxide/Co inerface due to interfacial electric field effect, which is different to Fert-Levy DMI at FM/NM interfaces. Finally, we demonstrate that DMI amplitude can be modulated using an electric field with efficiency factor comparable to that of the electric field control of perpendicular magnetic anisotropy in transition metal/oxide interfaces. These approaches of DMI controlling pave the way for skyrmions and domain wall motion-based spintronic applications.

\end{abstract}

\maketitle

The possibility to manipulate magnetization at nanoscale using the coupling between electron's spin and its motion (orbital angular momentum) has led to the emergence of a new research field named "spin-orbitronics". A fascinating example of the impact of the spin-orbit coupling (SOC) on the magnetization profile is the chiral spiral or skyrmion magnetic orders observed at the surface of magnetic ultrathin films \cite{Roessler,Muhlbauer,Yu, BlugelNP, Yoshida, Dupe, Simon,Bode,parkin1,BoulleNN}. Such spin configurations are driven by an additional term in the exchange interaction, namely Dzyaloshinskii-Moryia interaction (DMI) \cite{D,M,FertLevyDM, BlugelPRB, PtCoNiPRL2015, PyPtTiusanPRB2015, PtCoAlOThiavillePRB2015, Emilie,Berkeley,multilayer,neelDMI}, which arises from the presence of SOC and inversion symmetry breaking\cite{FertLevyDM,FertSky1}. Novel out-of-equilibrium spin transport phenomena also results in such structure, such as the "spin-orbit torques"(SOT) exerted on the magnetization when injecting a current, leading to fast current induced domain wall motion and magnetization reversal\cite{m3, miron2011NM,Ralph,experDM,parkin1,Andre}. This has led to new concept of magnetic memory, such as skyrmions-based memory, where the information is coded by nm scale magnetic skyrmions in nanotracks and manipulated using current pulses \cite{FertSky1,FertSky2,Nagaosa}. However, for the applications, there are still many issues to solve, for example in order to increase the stability of the chiral dependent domain wall (DW), the increasing of DMI amplitude is critical. Furthermore, the velocity of domain wall motion is also shown strongly depending on DMI amplitude. Therefore, searching large DMI is the key to realize stable skyrmions and chiral DW for spin-orbitronic devices.  

\begin{figure}
   \includegraphics[width=\textwidth]{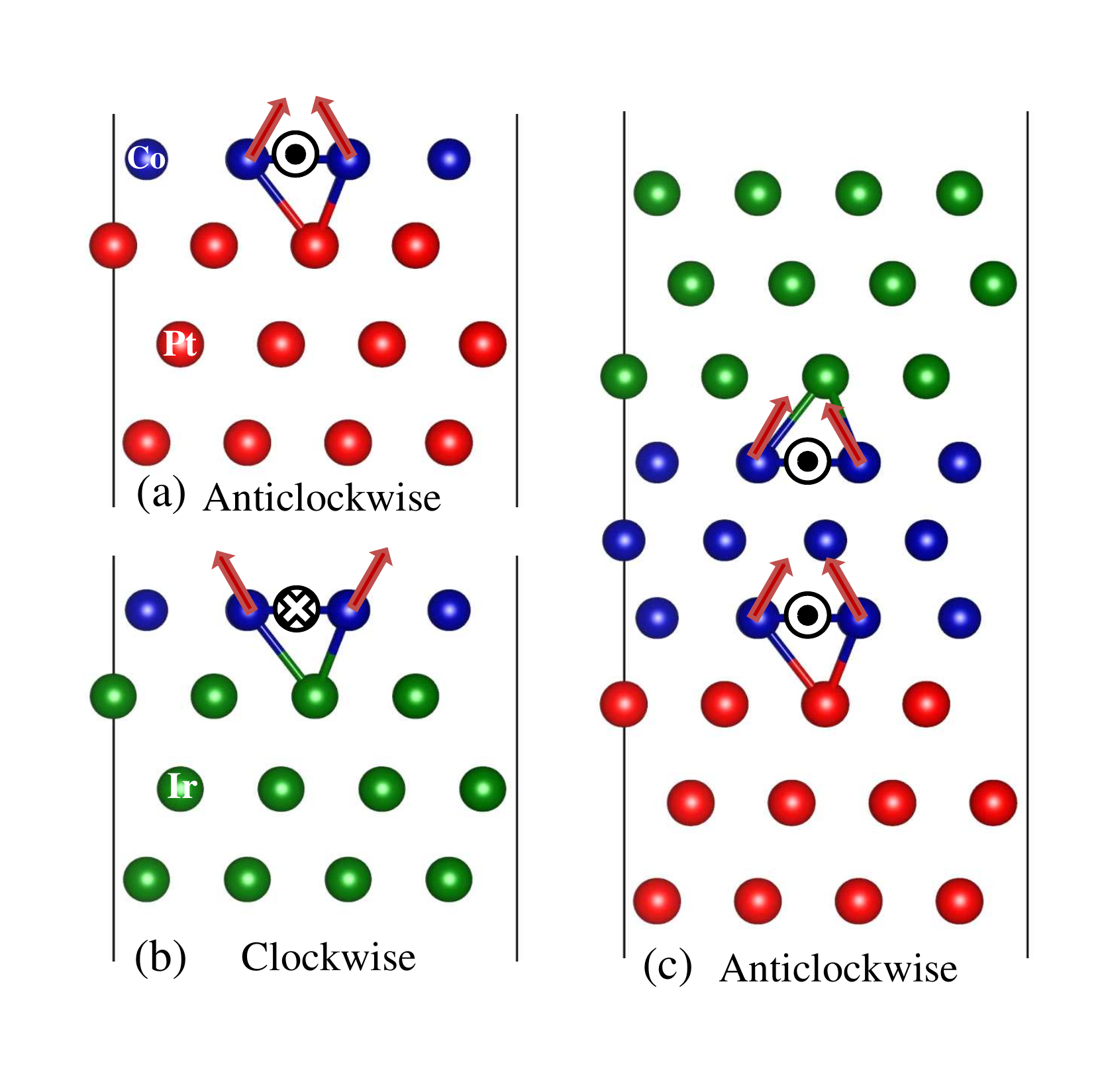}
   \caption{Schematic structures with anticlockwise DMI in Co/Pt bilayers (a), clockwise DMI in Co/Ir bilayers (b) and enhanced DMI in Ir/Co/Pt trilayers (c).}\label{fig1}
\end{figure}

In this Letter, we propose several approaches to enhance DMI in ultrathin magnetic films. First, we show that DMI can be magnified via multilayer stacking of FM and NM metals. Such a multilayer stacking approach is a powerful tool in spintronics to engineer device transport and magnetic properties~\cite{multilayer,Leeds,multilayerNC,FeNiPMA}. Here, in case of DMI enhancement, the key is to find required DMI chiralities for additive effects at successive interfaces. For example, in asymmetric trilayers of Ir/Co/Pt, where the DMI chirality at separated Co/Pt and Co/Ir interfaces is opposite as shown in Fig.~\ref{fig1}(a) and (b), respectively. Due to the inversion geometry stacking from Co/Ir to Ir/Co in forming Ir/Co/Pt trilayers, the sign of DMI at the interface of Ir/Co is reversed resulting in an overall enhanced anticlockwise DMI as schematiclly shown in Fig.~\ref{fig1}(c).
Next approach we propose to use oxidized layer, such as MgO, capping on top of Co/Pt bilayers, which is shown to efficiently enhance the DMI. Here, the DMI enhancement mechanism arising from MgO/Co interface is found from Rashba effect. Finally, we demonstrate that DMI can be efficiently tailored by applying an electric field (EF). This unveils the possibility to control DMI and perpendicular magnetic anisotropy (PMA) simultaneously via electric field which opens an efficient route towards EF-manipulation of magnetic skyrmions.

\begin{figure}
    \includegraphics[width=\textwidth]{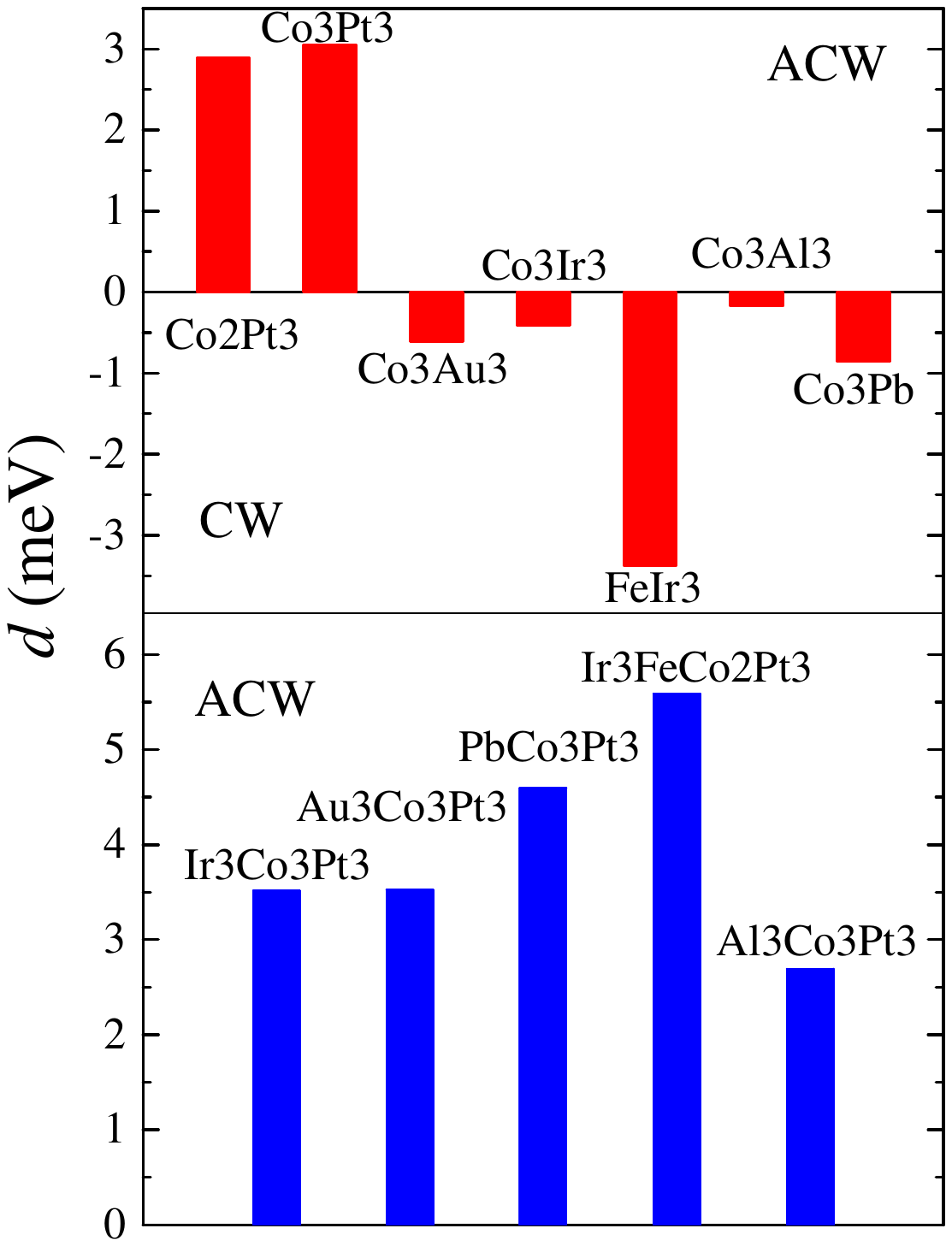}
   \caption{Calculated microscopic DMI, $d$, for bilayer structures (upper panel), and trilayers (lower panel).}\label{fig2}
 \end{figure} 
 
To calculate DMI, we employ the same approach as in our previous work~\cite{yang2015arXiv} based on the Vienna $ab$~$initio$ simulation package (VASP)~\cite{vaspPRB93,vaspPRB96}, which represent an adaptation to the case of layered structures of the method used for DMI calculations in bulk frustrated systems and insulating chiral-lattice magnets~\cite{methodPRB, methodPRL}.
The electron-core interactions are described by the projected augmented wave (PAW) method for the pseudopotentials with exchange correlation energy calculated within the generalized gradient approximation (GGA) of the Perdew-Burke-Ernzerhof (PBE) form~\cite{pbe}. 
The cutoff energies for the plane wave basis set used to expand the Kohn-Sham orbitals were chosen to be 320~eV and 500~eV for structures comprising only metals and those with oxygen, respectively. 
The Monkhorst-Pack scheme was used for the $\Gamma$-centered 4$\times$16$\times$1 k-point sampling. 

Our DMI strength parameter $d$ is derived from the energy difference between clockwise and anticlockwise magnetic spiral configurations by identifying this difference with what can be expected from an analytical expression of pair DMIs for the atoms in a single Co atomic layer.\cite{yang2015arXiv}. This $d$ given in eV per atom can be seen as the coefficient of pair DMIs concentrated in a single atomic layer and producing an equivalent effect~\cite{yang2015arXiv}. The calculation of $d$ has been performed in three steps. 
First, structural relaxations were performed until the forces become smaller than 0.001~eV/\AA~for determining the most stable interfacial geometries. 
Next, the Kohn-Sham equations were solved, with no SOC, to find out the charge distribution of the system's ground state. 
Finally, SOC was included and the self-consistent total energy of the system was determined as a function of the 
orientation of the magnetic moments which were controlled by using the constrained method implemented in VASP. 

\begin{figure}
  \includegraphics[width=\textwidth]{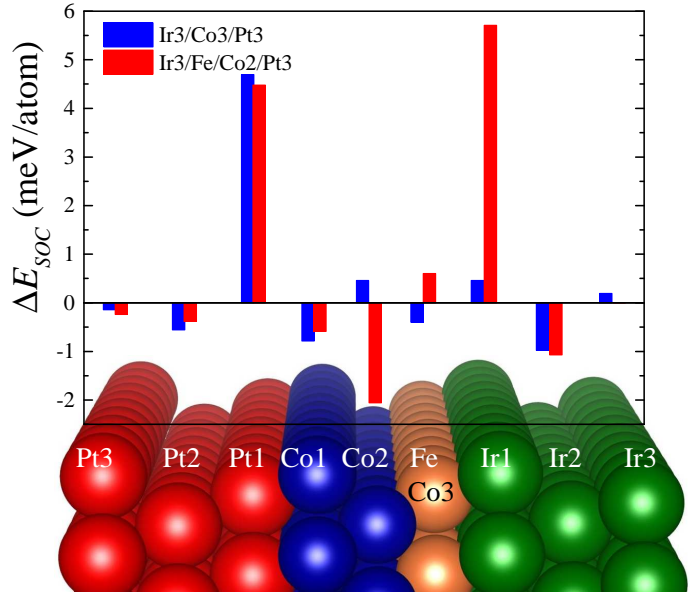}
   \caption{
Layer-resolved distribution of the energy difference between clockwise and anticlockwise chiralities for Ir3Co3Pt3 (blue) and IrFeCo2Pt3 (red) structures, respectively.
}\label{fig3}
 \end{figure}
 
In order to design metallic trilayers with efficiently enhanced DMI, we first systematically investigated DMI in FM/NM bilayers.
The calculated results are summarized in the upper panel of Fig.~\ref{fig2}. One can see that DMI chirality is anticlockwise (ACW) for Co/Pt bilayers while for Co on Au, Ir, Pb and Al substrates, DMI shows a clockwise (CW) chirality. We note that the DMI magnitude is much larger for Fe on top of Ir compared to Co on Ir case. 
Let us now combine two interfaces of opposite DMI chiralities, e.g. Co/Pt and Co/Ir shown in Fig.~\ref{fig1}(a) and (b), with reverse stacking of the second interface so that both interfaces in the resulting Ir/Co/Pt structure [Fig.~\ref{fig1}(c)] possess ACW chirality. As summarized in the bottom panel of Fig.~\ref{fig2}, the resulting DMI strength for NM/Co/Pt trilayers is approximately equal to the sum of DMI amplitudes of the two interfaces, i.e. Co/Pt and Co/NM, constituting the trilayer structure. Here, it demonstrates again that DMI is localized at interface.\cite{yang2015arXiv}

Due to small DMI amplitude at Co/NM interfaces in cases of NM=Al,Au,Ir, the DMI strength in NM/Co/Pt trilayers is moderately enhanced (\textless 15~\%) compared to that of Co/Pt bilayer, 3.05 meV/atom. 
However, in the case of Pb/Co/Pt trilayer, the DMI enhancement can reach up to 4.59 meV/atom. This large enhancement is due to the high DMI value at the Pb/Co interface (0.85 meV/atom with CW chirality), as compared to that of Co/Au and Co/Ir interfaces.
It is interesting to note that the total DMI of Pb/Co/Pt structure, 4.59 meV/atom, is larger than the sum of the DMI contributions from Co/Pt and Pb/Co bilayers, i.e. 3.05 meV/atom + 0.85 meV/atom = 3.90 meV/atom.  
This discrepancy is due to the DMI from 2nd and 3rd interface layers ~\cite{yang2015arXiv}, and the influence of the electronic structure, especially the shifting of Fermi level in trilayers compared to the corresponding bilayers. This deviation is also observed in other trilayers [Fig.~\ref{fig2}].

\begin{figure}
  \includegraphics[width=\textwidth]{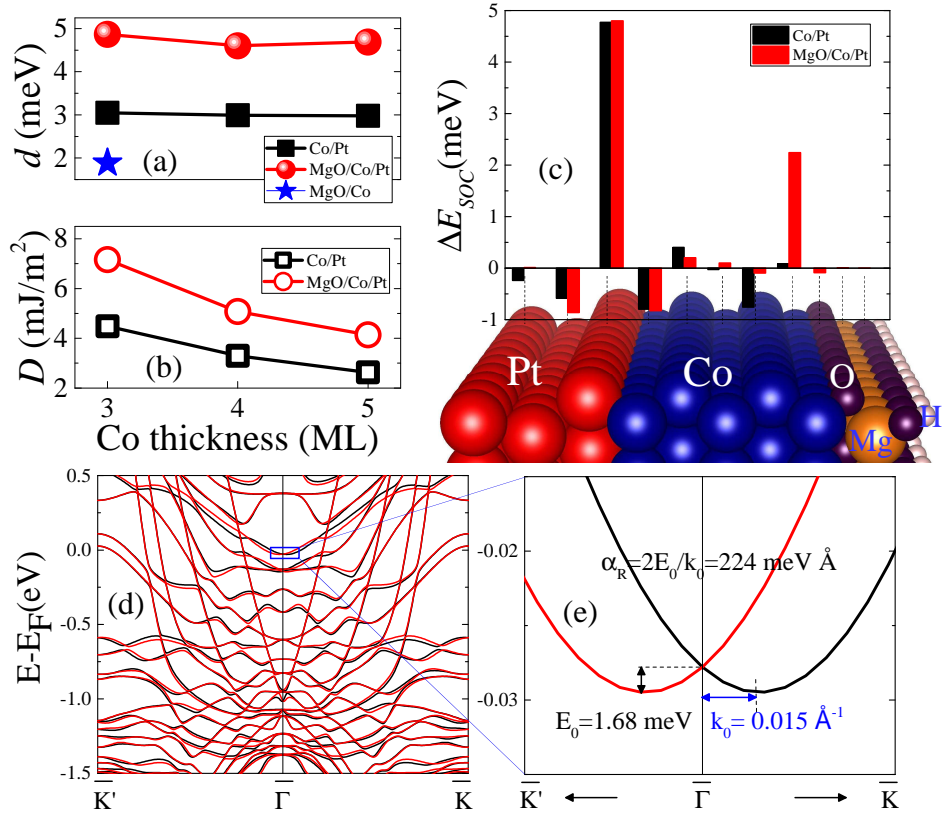}
   \caption{(a) $Ab-initio$ calculated DMI for microscopic, $d$, in Co/Pt (solid black squire) and MgO/Co/Pt (solid red ball) and MgO/Co (blue star) structures as a function of Co thickness. (b) $Ab-initio$ calculated micromagnetic DMI, $D$, in Co/Pt (empty black squire) and MgO/Co/Pt (red circle) as a function Co thickness. (c) Spin-orbit coupling energy difference,~$\bigtriangleup E_{SOC}$, in Co/Pt (black bars) and MgO/Co/Pt (red bars) between clockwise and anticlockwise spin spirals. (d) Band structure for MgO capped 3ML of Co when the magnetization is setting along $\langle$110$\rangle$ (black) and $\langle$\={1}\={1}0$\rangle$ (red) when SOC is switched on. (e) Zoom in of the band structure around $\bar{\Gamma}$ point to calculate Rashba coefficient.} \label{fig4}
\end{figure}
 
Next, we consider a four layer Ir/Fe/Co/Pt structure which can be seen as replacing one monolayer of Co in Ir/Co/Pt trilayer by Fe. This is particularly interesting since a single layer of Fe on Ir(111) possess very large DMI compared to that of Co on Ir [Fig.~\ref{fig2}]. The calculated DMI for this structure is shown in the bottom panel of Fig.~\ref{fig2}. One can see that there is a significant enhancement for DMI in Ir/Fe/Co/Pt compared to that in Ir/Co/Pt with amplitude up to 5.59 meV/atom which is 1.8 times larger compared to that of Co/Pt. 

In order to elucidate microscopic mechanisms of Fe induced DMI enhancement in this structure, we calculated the chirality-dependent SOC energy difference, $\triangle E_{SOC}$, for both Ir3/Co3/Pt3 and Ir3/Fe/Co2/Pt3 as shown in Fig.~\ref{fig3}. 
For the FM layer containing only three ML of Co (blue bars), the dominant SOC energy source $\triangle E_{SOC}$ is located at the first interfacial Pt layer, indicating large DMI in adjacent ferromagnetic Co1 layer.\cite{yang2015arXiv} Whiles, the $\triangle E_{SOC}$ at the Ir/Co interface is much smaller compared to that at the Co/Pt interface. However, when the adjacent Co to Ir layer is replaced by Fe, $\triangle E_{SOC}$ located at interfacial Ir layer is significantly increased (red bars) indicating a large DMI at Fe layer. 
Thus, Ir/Fe/Co/Pt structure is much more efficient than Ir/Co/Pt for providing high DMI values.

\begin{figure}
  \includegraphics[width=\textwidth]{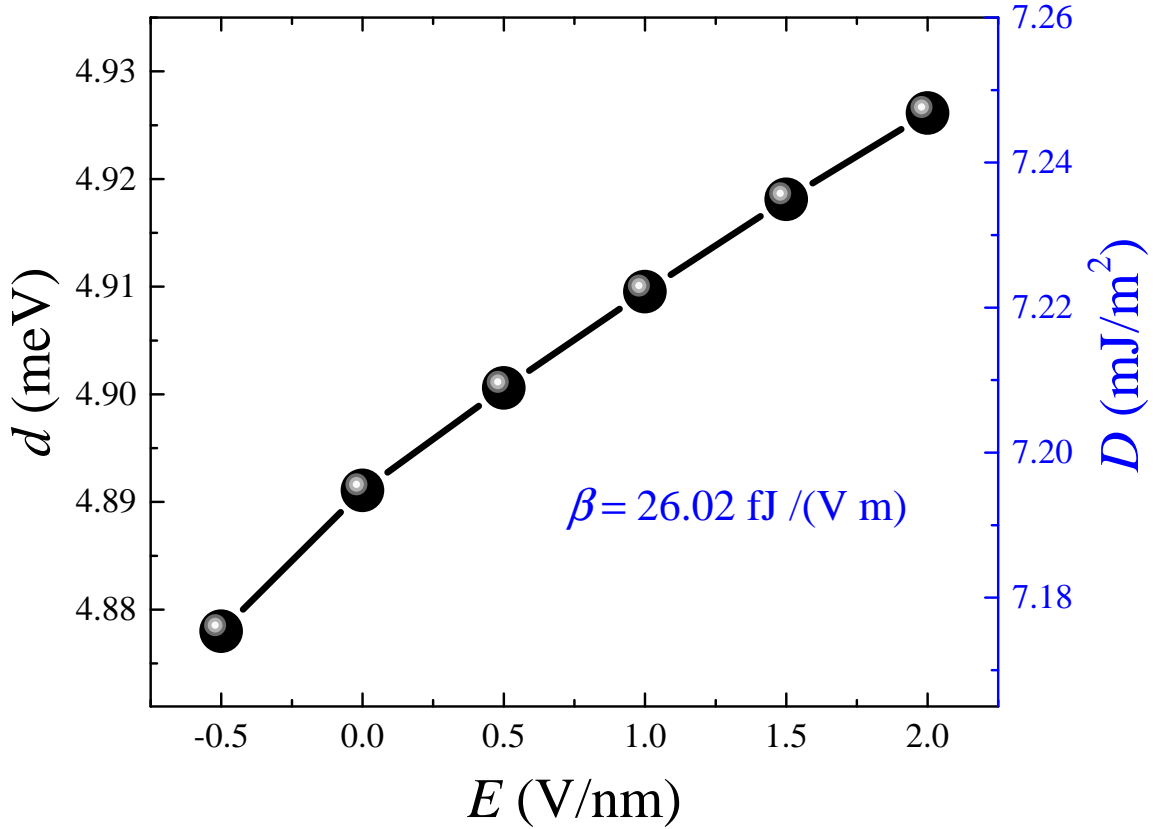}
   \caption{ Microscopic and micromagetic DMI for MgO/Co/Pt structure shown in Fig.~\ref{fig4} as a function of electric field applied along normal orientation of the surface. Positive electric field means the electric field is applied from insulator to metal. The slope $\beta $~ is indicated for micromagetic DMI.
}\label{fig5}
 \end{figure}
 
Another approach to enhance DMI is to add an insulator (I) capping layer on top of FM/NM to form I/FM/NM structure.  
These structures, notably AlOx/Co/Pt, MgO/Co/Pt or MgO/FeCo/Ta, are of major interest for the manipulation of magnetization~\cite{miron2011NM,m3,Ralph,experDM,ptcomgoAPL,neelDMI,BoulleNN} in domain walls, skyrmions or nanomagnets via spin-orbit torques, where DMI is known to play an essential role in the magnetization reversal~\cite{Andre,OBprl,Mikuszeit,Martinez}. In particular, DMI is expected to lead to a large decrease of the switching current in perpendicular magnetized nanomagnets~\cite{Mikuszeit}.
We consider the MgO/Co/Pt structure comprises 3 ML of Pt, 3 to 5 ML of Co, and MgO with surface passivated by hydrogen as shown in Fig.~\ref{fig4}(b). The results show that DMI in MgO/Co/Pt is much larger, about 1.6 times, compared to that of Co/Pt bilayers for all the Co thicknesses considered. It is shown that when vary the thickness of Co layer from 3-5 MLs in both Co/Pt and MgO/Co/Pt structures, microscopic DMI $d$ is not affected much suggesting that the Co/MgO interface affects almost equally to the electronic structure of the Co layer up to at least 5 atomic layers [shown in Fig.~\ref{fig4}(a)]. As for the micromagnetic DMI constant $D$ defined according to Eq.~4 in Ref.~\cite{yang2015arXiv} is expected to decrease with increasing Co thickness [Fig.~\ref{fig4}]. The strong enhancement of DMI in MgO/Co/Pt has been observed in experiments for 5 ML Co thickness case in ref.10, where the micromagnetic DMI from experiment and $ab~initio$ are 2.0 and 3.9 mJ/m$^2$, respectively.
In order to understand the mechanism of the DMI enhancement due to the MgO capping, we first calculated the DMI at MgO/Co interface, we see the DMI amplitude is about 1.88 meV/atom [Fig. 4(a)]. Next, we did the analysis of SOC energy difference between CW and ACW spin spirals for Co/Pt [black bars] and MgO/Co/Pt [red bars] in Fig.4(c). Very interestingly, one can see that at Co/Pt interface, the SOC energy source is located in the interfacial Pt layer, which is a typical Fert-Levy model Ref.~\cite{yang2015arXiv}, 
whereas at MgO/Co interface, the DMI is localized in the interfacial Co layer Ref.~\cite{BoulleNN}, and the SOC energy source also located in the same interfacial Co layer, which indicates a different mechanism. Actually, this DMI at MgO/Co interface is governed by the Rashba effect \cite{ZhangS, Bruno, Stiles}, where the DMI can be expressed as $d$=2$k_RA$. In which, $A$ is the exchange stiffness, $k_A$=$\dfrac{2\alpha_Rm_e}{\hbar^2}$ is determined by the Rashba coefficient, $\alpha_A$, and effective mass, $m_e$. The effective mass in Co was measured to be about 0.45 $m_0$ ~\cite{mass} and the exchange stiffness, $A$ was found to be about 15.5 $\times$ 10$^{-12}$ J/m \cite{exchange,vincentNN}. The Rashba coefficient, $\alpha_R$, can then be extracted from $\alpha_R$=2E$_0$/k$_0$, where E$_0$ is the Rashba splitting at the wave vector k$_0$. We calculated the Rashba splitting for the MgO capped 3ML of Co case by switching on SOC and putting the magnetization along $<$110$>$ and $<\bar{1}\bar{1}0>$~\cite{blugelRashba}, as shown in Figs. 4(d) and 4(e) with black and red curves, respectively. We used the band around Fermi level at the $\bar{\Gamma}$ point, as shown in Fig. 4(e), to estimate the Rashba-type DMI, where the splitting, E$_0$, is about 1.68 meV at k$_0$=0.015 \AA$^{-1}$, and the Rashba coefficient, $\alpha_R$, is found to be about 224 meV$\cdot$\AA. We found then k$_R$=26.6$\times$10$^{-3}$ \AA$^{-1}$ and therefore d=0.81 meV, which is smaller than the DFT calcualted DMI 1.88 meV. The reason for the smaller DMI value extracted from the Rashba effect can be ascribed to the fact that the Rashba-type DMI was calculated by using only one band close to Fermi level, whereas other bands further from Fermi level may also contribute to the total DMI.

Finally, we explore electric field control of DMI in Oxide/FM/NM structures. Electric field control of interfacial magnetism, in particularly of PMA has been an intense interest\cite{Givord, suzuki0, suzuki1}. However there is no first-principles study of electric field control of DMI even though it has been experimentally reported recently for the very small DMI case in MgO/Fe/Au multilayer~\cite{suzuki2}. Here we study the effect of an electric field on the DMI in MgO/Co/Pt trilayers. We used 3 ML of Co structure to calculate DMI as a function of external electric field $E$ (see Fig.~\ref{fig5}). The electric field is applied perpendicularly to the plane of interface with positive voltage pointing from insulator to metal. It is shown that for both microscopic DMI $d$ and micromagnetic $D$, are approximately linear-increasing as a function of $E$ ~[Fig.~\ref{fig5}]. Similarly to the electric field control of the PMA, the efficiency of the EF control of the DMI can be characterized by the slope of the curve $\beta$ defined as a ratio of the DMI change to $E$, which is found to be equal to 26.02 fJ/(Vm). Interestingly, this parameter is comparable to the slope in electric field control of PMA for Fe(Co)/MgO structures~\cite{Fatima, Suzuki3,suzuki2}, which unveils the possibility of simultaneous tuning of both PMA and DMI within the same range suggesting a route towards an efficient way for controlling skyrmions since they depend on both the DMI and PMA.

In conclusion, we proposed three approaches for the efficient tuning of Dzyaloshinskii-Moriya interaction (DMI). The first one is to use NM/FM/Pt trilayers with inverse stacking of FM/Pt and FM/NM structures with opposite DMI chiralities. This allows the enhancement of DMI up to 50\% as compared to the corresponding FM/Pt bilayers. Moreover, we demonstrated that in case of Ir/Fe/Co/Pt multilayers a giant DMI values up to 5.5 meV/atom can be achieved, which is almost twice of that for Co/Pt bilayers. The second approach is to cap Co/Pt structure with an oxidized layer, which can cause a dramatic DMI enhancement due to the Rashba type DMI. Finally, we demonstrated that DMI can be controlled by the application of an electric field in MgO/Co/Pt structure and showed that its efficiency is comparable to E-field control for PMA. These three very efficient approaches pave the way for engineering giant DMI for spintronic applications.

\acknowledgement
The authors thank A. Thiaville, I. M. Miron, G. Gaudin, L. Buda-Prejbeanu, S. A. Nikolaev, S. Rohart, S. Pizzini, J. Vogel, N. Reyren and C. Moreau-Luchaire for fruitful discussions. This work was supported by ANR ESPERADO, SOSPIN and ULTRASKY.


\end{document}